\documentclass[manuscript]{aastex}

\shorttitle{Collapsed Cores in Globular Clusters}
\shortauthors{Djorgovski et al.}

\begin{document}


\title{Analytic Core Envelope Models for the Crab and the Vela Pulsars}


\author{P. S. Negi}
\affil{Department of Physics, Kumaun  University,
    Nainital 263 002, India}

\email{psneginainital63@gmail.com; psnegi\_nainital@yahoo.com}





\begin{abstract}
We study analytic core-envelope models obtained in Negi et al. (1989) under slow rotation. We have regarded in the present study,  the lower bound on the estimate of moment of inertia of the Crab pulsar, $I_{\rm Crab,45} \geq 2$ (where $I_{45}=I/10^{45}\rm g{cm}^2$) obtained by Gunn and Ostriker (1969)  as a round off value of the recently estimated  value of  $I_{\rm Crab,45} \geq 1.93$ (Bejger \& Haensel 2003) for the Crab pulsar. If this value of lower bound is combined with the other observational constraint obtained for the Crab pulsar (Crawford \&  Demiansky 2003), $G_h = I_{\rm core}/I_{\rm total} \geq 0.7$ ( where $G_h$ is called the glitch healing parameter and represents the fractional moment of inertia of the core component  in the starquake mechanism of glitch generation), the models yield the mass, $M$, and surface redshift, $z_a$, for the Crab pulsar in the range, $M = 1.79M_\odot - 1.88M_\odot$; $z_a = 0.374 - 0.393$ ($I_{45} = 2$) for an assigned value of the surface density, $E_a = 2\times 10^{14}\rm g{cm}^{-3}$ (like, Brecher and Caporaso 1976). This assigned value of surface density, in fact, is an outcome of the first observational constraint imposed on our models that further yields the mass   $M = 1.96M_\odot $ and surface redshift $z_a=$ 0.414 ($I_{45}= 2$) for the values of  $G_h \approx 0.12$, which actually belongs to  the observed `central'  weighted mean value for the Vela pulsar. These values of mass and surface redshift predict the energy of a gravitationally redshifted electron-positron annihilation line, $E (\rm {MeV}) = 0.511/(1+z_a)$ (Lindblom 1984) in the range about 0.396 - 401 MeV from the Crab and about 0.389 MeV from the Vela pulsar. The evidence of a line feature at about 0.40MeV from the Crab pulsar (Leventhal et al. 1977) agrees quite well with the finding of this study.
\end{abstract}




\section{Introduction}

The estimation of mass and moment of inertia of slowly rotating pulsars like the Crab and the Vela is a topic of large interest
in the field of Astrophysics. The mass of a static (non rotating) Neutron Star (NS) depends on the Equation of State (EOS)
or the density distribution considered inside the spherical configuration which may be constrained on the basis of observationally 
measured values of NS masses. However for slowly rotating configurations, the matter distribution inside the structure is affected by 
the General Relativistic effects like the dragging of the local inertial frame. Haensel (1990) showed that the maximum moment of 
inertia of theoretical models of NSs may increase by a factor of seven as compared to the maximum mass of a theoretical model which increases
only by a factor of two from going to softer to the stiffest EOSs.

In earlier attempts, by assuming a certain mechanism for the loss of rotational energy of the Crab pulsar, Cohen \& Cameron (1971) estimated the 
moment of inertia of the Crab pulsar, $I_{\rm Crab,44} \geq 1.8$, if only electrons are assumed to be involved in the synchrotron radiation
(Baldwin 1971). If, on the other hand, protons are also considered to take part  in the synchrotron radiation, the Crab pulsar yields a moment of inertia
, $I_{\rm Crab,44} \geq 4$ (Goldreich \& Julian 1969). Yet another mechanism (Gunn \& Ostriker 1969) leads  $I_{\rm Crab,45} \geq 2$  for the Crab pulsar.
By using the recent observational data, Begjer \& Haensel (2003; and references therein) obtained a lower bound on the moment of inertia of the Crab pulsar $I_{\rm Crab,45} \geq 1.93$ (for the
Crab nebula mass $M_{\rm nebula} = 4.6M_\odot$ in a `realistic' time dependent acceleration  model for the nebula) . Thus
the value obtained by Gunn \& Ostriker (1969) may be regarded as good as (round off) the value obtained by Begjer \& Haensel (2003) for the lower bound of the moment of inertia of the Crab pulsar.
The  lower bound on the moment of inertia of the Crab pulsar obtained by Begjer \& Haensel (2003) yields the mass of the Crab pulsar $M > 1.5 M_\odot$ and radius $a = 11 - 15 \rm {km}$ among the thirty EOSs of dense nuclear matter for NS models.

In an another study, Crawford \& Demiansky (2003) have shown that from the 21 measurements of Crab glitches, a weighted mean of the values yields   $G_h  \geq 0.72 \pm 0.05$. Thus the range  $G_h  \geq 0.7$  encompasses the observed distribution for the Crab pulsar for which  the `realistic' mass range for Crab pulsar, $M = 1.4\pm 0.2M_\odot$, yields by the seven representative EOSs for NS models. However, the 11 measurement for Vela glitches yields a weighted mean of $G_h = 0.12\pm 0.07$. Thus for much lower values of Vela glitches in the range  $G_h  \leq 0.2$, their models yields a unrealistic low mass range for the Vela pulsar, $M \leq 0.5M_\odot$. Therefore the study of  Crawford \& Demiansky (2003) had concluded that the starquake is the viable mechanism for glitch generation in the Crab and the vortex unpinning (the another mechanism) is suitable for the Vela pulsar which can be understood on evolutionary grounds.

In recent studies, Zhou et al (2014) discussed the starquake model for the Vela pulsar on the basis of a solid quark star model. Lai et al (2018) have proposed a strangeon star model  (i.e., the solidification of the star takes place during cooling) and studied the behaviors of glitches (as result of starquake) without significant energy release, including the Crab and the Vela pulsars. Haskell et al (2018) have constructed a model which naturally produces large glitches as observed in the Crab pulsar (including the glitch of the year 2017). A comparison of their study with observations of the large glitches in the Crab and the Vela suggested the crustal origin for the Crab glitches, but an outer core contribution for the Vela glitches. By using NS observations, Steiner et al (2015) obtained a crustal fraction of the moment of inertia as large as 10\% for a mass $M = 1.4M_\odot$ to explain the glitches in the Vela pulsar even with a large amount of superfluid `entrainment' (Andersson et al 2012; Chamel 2013). Delsate et al (2016) have calculated the crustal moment of inertia of glitching pulsars for different unified dense matter EOSs in order to explain the large glitches observed  in the case of the Vela pulsar.

Earlier, by considering the starquake as a feasible mechanism of glitch generation the author (Negi 2011) constructed NS models  which have been successful to  satisfy the observed glitch healing parameter for both the pulsars, the Crab and the Vela, simultaneously.
By imposing recent  constraint on the moment of inertia of the Crab pulsar mentioned above, the author co-related the mass of the Crab and the Vela pulsars (on the basis of observed values of $G_h$ of these pulsars) and obtained the lower bound as 2.11$M_\odot(G_h \approx 0.7$) for the Crab and upper bound as 0.98$M_\odot(G_h \approx 0.2$) for the  Vela pulsar. The present study continues to deal with the construction of such NS models which  can satisfy  both of the observational constraints mentioned above together with the success that  the `realistic' mass range for the Vela pulsar as well as that of the Crab pulsar may be obtained. Since for realistic NS models it is expected that the (energy) density in the core region varies smoothly as compared to the outer region [we shall call this region as `envelope' of the star, that is, right from the core-envelope boundary to the surface of the star (like Negi 2011, 2006)] where the variation of density becomes much faster. We have, therefore, considered in the present study, the core-envelope models presented in Negi et al. (1989) in which the density in the core is governed by the smooth variation of density (Tolman's type VII solution) and density in the envelope is described by the fastest (inverse square) variation of density (Tolman's type VI solution) and this model also fulfills various criteria of physically realistic structures (Durgapal et al. 1984). In the model described by Negi et al (1989), the density at the surface ($E_a$) remains finite where the pressure becomes vanishingly small (i.e., the matter represents a self bound state at the surface).   Because the core-envelope boundary  plays a crucial and important role for obtaining glitch healing parameters of NS models considered in any study, the distinctive feature of the present study lies in the fact that unlike other models available in the literature in which the core-envelope boundary is chosen in somewhat  arbitrarily manner (See e.g. Shapiro \& Teukolsky 1983; Datta \& Alpar 1993), the boundary of the core-envelope models considered in Negi et al. (1989) was obtained analytically and  in an appropriate manner by matching of  all the four variables viz. pressure ($P$), energy density ($E$) and both of the metric parameters ($\nu$ and $\lambda$) at the core- envelope boundary without recourse to any computational method.

Furthermore,  the density at the  core-envelope boundary, $E_b$, of the present models yields in the range $E_{\rm ave} \leq E_b \leq 4E_{\rm ave}$ (where $E_{\rm ave} =  2\times 10^{14}\rm g{cm}^{-3}$ represent the average nuclear matter density) which  is the consequence of  the  first observational constraint ( $I_{\rm Crab,45} \geq 2$) imposed on our models. This density range is comparable with the corresponding density  range obtained by Kalogera \& Baym (1996) by using realistic EOS (Wringa et al 1988 [WFF88])  in the envelope and the extreme causal EOS in the core region of NS models. It is well known that the WFF88 EOS has been widely used in the literature for constructing the realistic NS models (see, e.g. Kalogera and Baym 1996; Crawford and Demianski 2003; and references therein).  Furthermore, a comparison and consistency regarding the total mass, total radius, central energy- density and the total moment of inertia of the Vela pulsar obtained on the basis of the present study with those of the  study carried out by Li et al (2016) on the basis of so called BCPM ( Barcelona – Catania –  Paris –  Madrid) EOS (Sharma et al. 2015) constructed from modern microscopic  BHF (Brueckner –  Hartree –  Fock) calculations (Baldo 1999) is discussed in Sec 3.
 Sect.2 of the present study deals with the relevant equations governing slow rotation of the analytic core-envelope models (Negi et al.1989). Results of the calculations and an application of the models to the Crab and the Vela pulsars are presented in Sect.3.  Sect 4  summarizes  the main  results of the present study.

\section{Equations Governing Slow Rotation of Spherical Configuration}

The metric corresponding to a static and spherically symmetric mass distribution is given by
\begin{equation}
ds^2 = e^\nu dt^2 - e^\lambda dr^2 - r^2(d\theta^2 +{\rm sin}^2\theta d\phi^2)
\end{equation}
where $G = c = 1$ (we are using geometrized units) and $\nu$ and $\lambda$ are functions of `r' alone.
The relevant equations governing the core ($0\leq r \leq b$) and  the envelope ($b\leq r \leq a$) are described respectively by Tolman's type VII and VI solutions of the metric (Eq.1) and are available in 
Negi et al.(1989). However, some relations relevant to the present study are redefined in the following (see Negi et al. 1989):

$u \equiv M/a$ is called the compactness parameter which is defined as the mass to size (radius) ratio of entire configuration; where
 the mass, $M = \int_{0}^{a} 4\pi E r^2dr$; and $y = r/a$ is called the radial coordinate measured in units of configuration size.

$Q(\equiv K^2/a^2)$ is defined as the compressibility factor, $K$ is a constant appearing in the Tolman's type VII solution. The matching of various parameters at the core-envelope boundary yields

$b^2/a^2Q = 5/6$, thus $(b/a)$ represents the boundary, $y_b$, of the core-envelope models and `b' represents the core radius. The relation between central to boundary density, $E_0/E_b$, and central to surface density are given by:
$(E_0/E_b) = 6$; $(E_0/E_a) = 7.2/Q$.

For slowly rotating spherical objects like the Crab and the Vela pulsars (rotation velocity about 188 and 70 rad sec$^{-1}$ respectively), the macroscopic parameters of the stars are affected by first-order rotation effects. Since the Lense-Thirring frame dragging effect is a first-order effect which turns out to be about 1-2 \%  for the Crab and the Vela pulsar. This effect is taken into consideration for calculating the moment of inertia of the models in the following manner [ note that the effects like deformation from spherical symmetry and mass shifts represent second order effects which are significant only for the millisecond pulsars and for the Crab and the Vela pulsars the second order effects turn out to be about 10$^{-4}$ or even lower(see, e.g. Arnet and Bowers 1977; Crawford and Demianski 2003). Therefore these effects can be safely ignored in the present study which deals with the macroscopic parameters of slowly rotating configurations]:

For slowly rotating structures a perturbation solution for the metric (Eq. 1) yields (Borner 1973; Chandrasekhar \& Miller 1974; Irvine 1978)
\begin{equation}
(d/dr)[F(d\chi/dr)] = G \chi
\end{equation}
where
\begin{equation}
F = e^{-(\nu+\lambda)/2}r^4
\end{equation}
\begin{equation}
G = 16\pi(P + E) e^{(\lambda - \nu)/2}r^4
\end{equation}
\begin{equation}
\chi = \omega - \Omega
\end{equation}
 $\omega$ being the angular velocity of structure and $\Omega$ the drag of the local inertial tetrad (Hartle \& Sharp 1965; Hartle 1967); known as Lense-Thirring effect.

Substituting $F(d\chi/dr) = \phi$ in Eq. (2) we get
\begin{equation}
d\chi/dr = \phi/F
\end{equation}
and
\begin{equation}
d\phi/dr = G\chi
\end{equation}
Substituting $r = ay$, $ d r = a dy$  in Eqs. (6) and (7) we get
$
d\chi/dy = (a\phi/F) = (a\phi/a^4f)
$
and
$(d\phi/dy) = G\chi a = a^2g\chi a
$
with $f =  e^{-(\nu+\lambda)/2}y^4$ and $g = 2(8\pi Pa^2 + 8\pi Ea^2)e^\lambda f$
or
\begin{equation}
[d/dy](\phi/a^3) =  g\chi
\end{equation}
and
\begin{equation}
d\chi/dy = (\phi/a^3)/f
\end{equation}
Substituting $\phi/a^3 = \psi$ in Eqs. (8) and (9) we have
\begin{equation}
d\psi/dy = g\chi
\end{equation}
\begin{equation}
d\chi/dy = \psi/f
\end{equation}
Eqs.(10) and (11) provide a set of two first order coupled differential equations which may be solved numerically by using the standard Runge - Kutta method with boundary conditions
\begin{equation}
\chi_{y=0} = 1; (d\chi/dy)_{y=0} = 0
\end{equation}
Integrating from the  centre ($y = 0$) to the surface ($y = 1$, i.e. $r = a$ and $P = 0$) of the configuration, we find that at the surface
\begin{equation}
\omega = \chi_a + (\phi_a/3a^3) = \chi_a + (\psi_a/3)
\end{equation}
Drag is given by the equation
\begin{equation}
\Omega = \omega - \chi; \rm {or} (\Omega/\omega) = 1 - (\chi/\omega)
\end{equation}
We define central drag as
\begin{equation}
 (\Omega/\omega)_0 = 1 - (1/\omega); \chi_0 = 1
\end{equation}
Thus the surface drag is given by
\begin{equation}
 (\Omega/\omega)_a = 1 - (\chi_a/\omega)
\end{equation}
The moment of Inertia, $I$, of the configuration is given by
\begin{equation}
  I = (\phi_a/6\omega) = (\psi_a a^3/6\omega)
\end{equation}



\section{Results and application of the models to the  Crab and the Vela Pulsars}
The calculations are performed for different values of $u$ in the range $0.1 \leq u \leq 0.25$ for assorted values of $Q$ ($0.1 \leq Q \leq 1.2$). it may be re-iterated that the maximum permissible value of $Q$ is 1.2 for which $(b/a) = 1$, i.e. the whole configuration pertains to just the core solution. For $Q = 0$ , we obtain $(b/a) = 0$ , that is there is no core and the whole configuration belongs to the envelope solution only.

Figure 1 shows the plot of moment of inertia $I$ in units of $a^3$ (`a' being the radius of the configuration) vs. $Q$ for different assigned values of $u$ ($0.1 \leq u \leq 0.24999$). Figure 2 is the plot of  moment of inertia in $\rm g {cm}^2$ (in $\rm { log}_{10}$) with changing $u$  of NS models for different values of $Q$ ($0.1 \leq Q \leq 1.1999)$. The surface density, $E_a$, of these models is assigned as the average nuclear matter density $2\times 10^{14}\rm g{cm}^{-3}$ (like, Brecher \& Caporosso 1976). Each of the horizontal lines drawn on  this plot show the lower bound on moment of inertia of the Crab pulsar obtained by Cohen \& Cameron (1971), Goldreich \& Julian (1969) and Gunn \& Ostriker (1969)[The lower bound on moment of inertia obtained recently by Bejger \& Haensel (2003) is considered in the present study as good as the value (round off) obtained by Gunn \& Ostriker (1969)] . Corresponding to these values of moment of inertia, the mass ($M$), radius ($a$), glitch healing parameter ($G_h$) and the surface redshift ($z_a$) of the  pulsars based on these core-envelope models are given in Table 1.

Figures 3 - 7 sketch the variation of relative drag ($\Omega/\omega$) vs. $y$, that is the change in drag from centre ($y=0$) to the surface ($y=1$) of the structures (for $u = 0.10, 0.15, 0.20,  0.24999$) under consideration for different values of $Q$ from 0.1 to 1.1999. It is seen that the the central drag $(\Omega/\omega)_0$ decreases with increasing $Q$ for all values of $u$. The surface drag  $(\Omega/\omega)_1$ remains almost same for all the values of $Q$, for $u =$ 0.10 to 0.24999.


\section{Conclusions}
The core-envelope models described in this paper obtain not only the lower bound on the moment of inertia of the Crab pulsar $I_{\rm Crab,45} \simeq 2$ on the basis of the  recent observational constraint on the Crab nebula mass $M_{\rm nebula} = 4.6M_\odot$ ( in a `realistic' time dependent acceleration  model for the nebula) but they can also satisfy the other observational constraint on the glitch healing parameter for the  Crab pulsar $G_h \geq 0.7$. For this value of moment of inertia and glitch healing parameters, the mass, $M$, and surface redshift, $z_a$, of the pulsar is obtained in the range $1.790 - 1.876 M_\odot$ and 0.374 - 0.393 respectively. These values of surface redshifts predict an electron- positron annihilation line in the range about 0.396 to 0.401 MeV. The observation of a line feature at about 0.40 MeV from the Crab pulsar by Leventhal et al. (1977) agrees quite well with the finding of this study. If, however the moment of inertia of the pulsar slightly exceeds the lower bound $I_{\rm Crab,45} = 2$, the mass and the surface redshift of the Crab pulsar also exceeds slightly upto the values of $ 1.920 - 1.964  M_\odot$ and 0.404 - 0.414 ($G_h \geq 0.7$) respectively.

The remarkable feature of this core-envelope model is that  it can also satisfy the much lower values of glitch healing parameters, $G_h \leq 0.35$, for the moment of inertia of the structure, $I_{\rm 45} \simeq 2$  for a mass, $M$, in the range 1.920 - 1.964 $M_\odot$ and the  surface redshift, $z_a$, in the range 0.404 - 0.414. For $G_h \approx 0.12$ (which is  the observed `central'  weighted mean value for the Vela pulsar) the mass and the surface redshift correspond to the values 1.964$M_\odot$ and 0.414 respectively. Since the much lower values of glitch healing parameters  observed for the Vela pulsar,  $G_h \leq 0.2$,  predicts much lower values of mass for the pulsar, $M \leq 0.5M\odot$ in the conventional models of NSs discussed in the literature, this study, on the other hand, shows that the mass and compactness of the Vela pulsar may be some what higher than that of the Crab pulsar  if the moment of inertia of Vela pulsar belongs to a value about $2 \times 10^{45} {\rm g cm^2}$.

The moment of inertia, mass and surface redshift predicted by this core-envelope model (Table 1) for  the Crab pulsar on the basis of the study of Cohen \& Cameron (1971) and Goldreich \& Julian (1969) are of the  academic interest only because of the recent estimation of the moment of inertia of the Crab pulsar. However, these models may become important for the Vela pulsar, if the forthcoming studies estimate the moment of inertia of the Vela pulsar also in the range provided by  Cohen \& Cameron (1971) and Goldreich \& Julian (1969) for the Crab pulsar.

The values of the energy-density at the core-envelope boundary, central energy-density, mass, radius and total moment of inertia of the models obtained in this study may be compared with those obtained in the literature by using realistic EOSs. For the values of  $I_{\rm {total},45} \geq 2$ the density at the core-envelope boundary, $E_b$, varies from $E_{\rm ave}$ to $4E_{\rm ave}$  which is comparable with the corresponding density range varies from $E_{\rm nm}$ to $4E_{\rm nm}$ (where  $E_{\rm nm} = 2.7 \times 10^{14}\rm gcm^{-3}$ represents the nuclear matter saturation density) obtained by Kalogera and Baym (1996)  by using WFF88 EOS in the envelope and the extreme causal, $\rm dP/\rm dE = 1$(in geometrized units), EOS in the core region of NS models. For the said values of the moment of inertia for our models the central energy- density, $E_0$ varies in the range $1.2 - 4.8\times 10^{15}\rm gcm^{-3}$ as shown in Table 1 which is also comparable with the corresponding range of the  values, $\simeq 8.8 \times 10^{14}\rm gcm^{-3}$ - $4.0 \times 10^{15}\rm gcm^{-3}$ obtained in the study of  Kalogera and Baym (1996). In a recent study, Li et al (2016) have predicted the internal structure of Vela pulsar on the basis of modern  BCPM  EOS (Sharma et al 2015) consided in the core and inner crust (equivalent to envelope in our model) was computed by using self-consistent Thomas– Fermi calculations with the BCPM functional in different Wigner Seitz (WS)  configurations, where the low-density neutron gas and the bulk matter of the high-density nuclear structures are given by the same microscopic BHF calculation. They have obtained the  total moment of inertia of the Vela pulsar about $ 2.16 - 2.22 \times 10^{45} {\rm g cm^2}$ for the assigned values of masses $M = 1.9 - 1.98M_\odot$ respectively. The total radius and the  central energy-density corresponding to the said masses are obtained as $10.8 - 10.1 \rm {km}$ and $1.45 - 1.88 \times 10^{15}\rm gcm^{-3}$ respectively. Thus except the central energy-density which is found  lower than that obtained in the present study, the other parameters corresponding to the Vela pulsar mentioned in the last two sentences are found in good agreement with the results obtained in the present study ( Table 1).

In summary, a comparison of the results obtained in  present core-envelope  analytic model  with those of the results obtained in the literature on the basis of various realistic EOSs for the core and the envelope regions reveals  that the analytic models may be completely analogous to the models represented by EOSs, provided they conform the physically realistic behavior. The present study, for example, indicates that instead of using BCPM/extreme causal  EOS in the core of the NS models (referenced above), one can employ Tolman's type VII solution in the core and  the  envelope of the models may be replaced by Tolma's type VI solution (or any other suitable analytic solution available in the literature). The elegance of using  analytic solution in the NS models lies in the fact that the various physical parameters like - pressure, energy-density, and both of the metric parameters ($\nu$ and $\lambda$)  have a direct dependence on the radial co-ordinate, $r$, in the interior which is not available in the NS models composed of EOSs. Thus the internal structure of NSs can be explored in a  more simple and elegant manner on the basis of analytic models.




\begin{acknowledgments}

We are grateful to Dr.P. K. Mishra and G. Punetha for providing LaTeX packege. The anonymous Referee is acknowledged for his valuable comments and suggestions which helped us to improve the present paper.

\end{acknowledgments}

\begin{figure}
\caption{Moment of inertia,$I$, of the core-envelope models in units of $a^3$ vs. $Q$ for different values of the compactness parameter $u$ in the range  0.10 - 0.24999.}
\includegraphics[angle=0,scale=1.0]{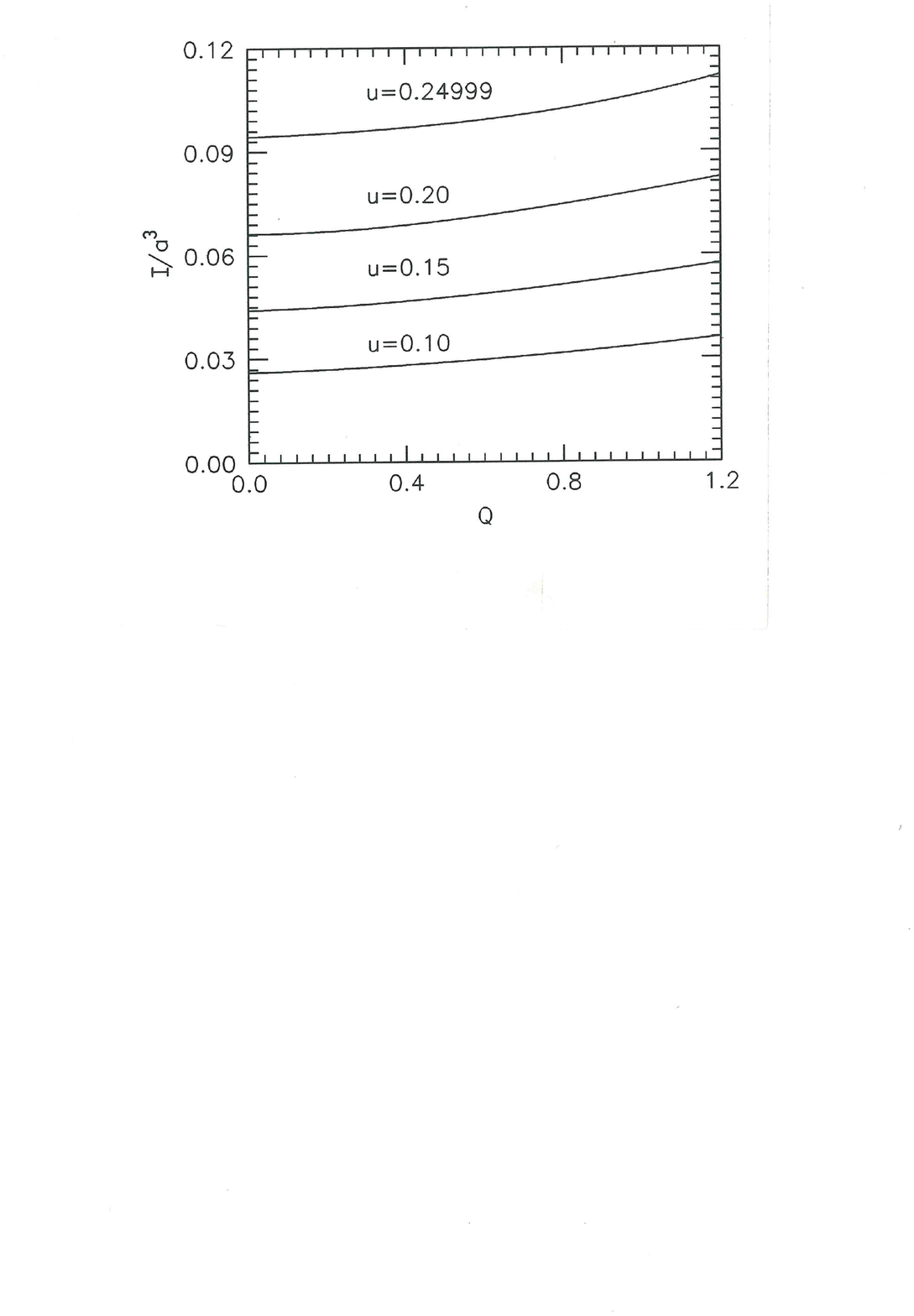}

\end{figure}



\begin{figure}
\caption{ The plot of  moment of inertia in $\rm g {cm}^2$ (in $\rm { log}_{10}$) vs.  $u$  for core-envelope models corresponding to different values of $Q$ ($0.1 \leq Q \leq 1.1999)$. The surface density, $E_a$, of these models is assigned as $2\times 10^{14}\rm g{cm}^{-3}$ (like, Brecher \& Caporosso 1976). Each of the horizontal lines drawn on  this plot show the lower bound on moment of inertia of the Crab pulsar obtained by Cohen \& Cameron (1971), Goldreich \& Julian (1969) and Gunn \& Ostriker (1969)[The lower bound on moment of inertia obtained recently by Bejger \& Haensel (2003) is considered in the present study as good as the value (round off) obtained by Gunn \& Ostriker (1969)]}
\includegraphics[angle=0,scale=1.0]{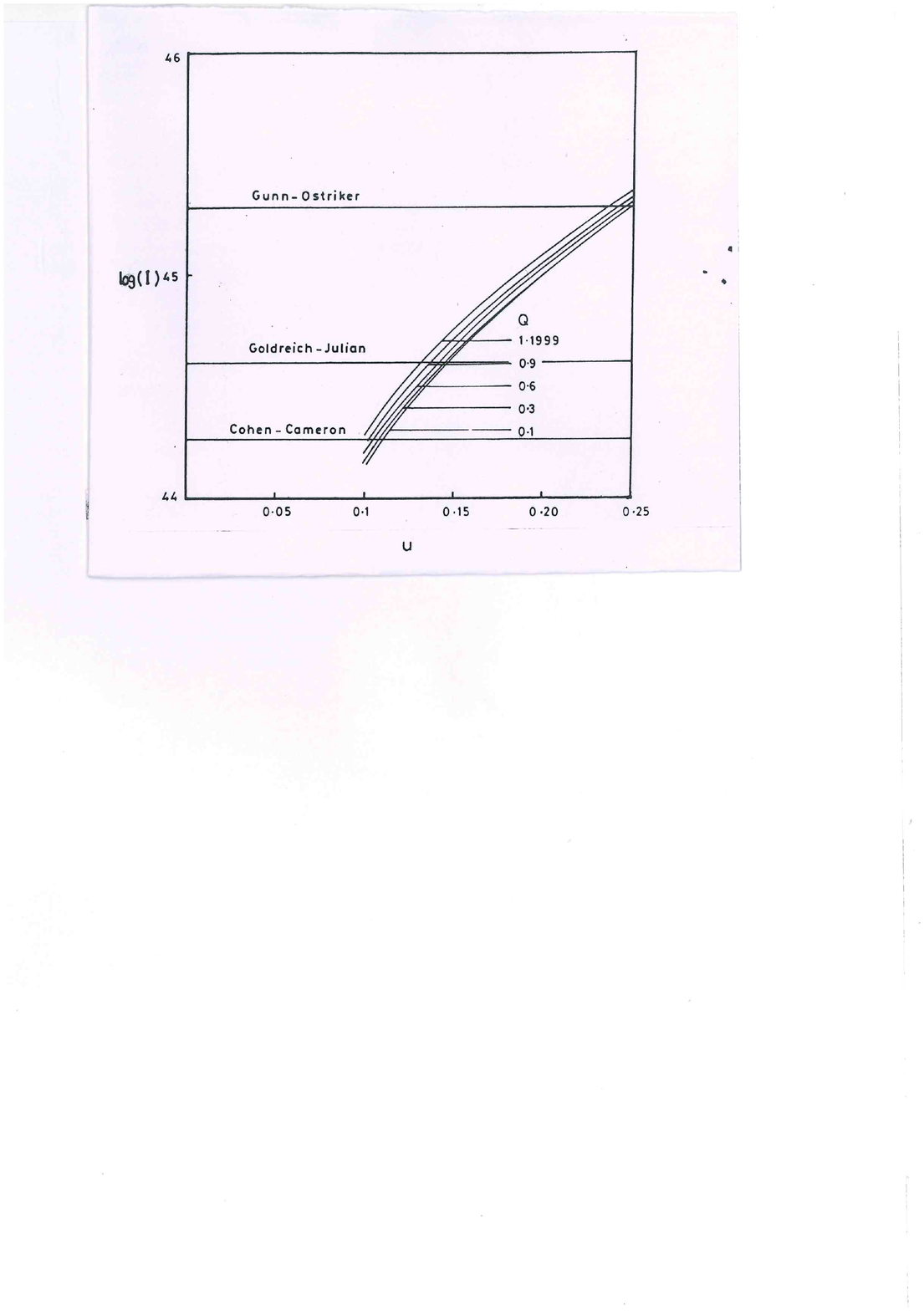}

\end{figure}


\begin{figure}
\caption{ The variation of relative drag ($\Omega/\omega$) vs. $y$, that is the change in drag from centre ($y=0$) to the surface ($y=1$) of the structures ($u = 0.10, 0.15, 0.20, 0.24999$) for the value of $Q = 0.1$.}
\includegraphics[angle=0,scale=1.0]{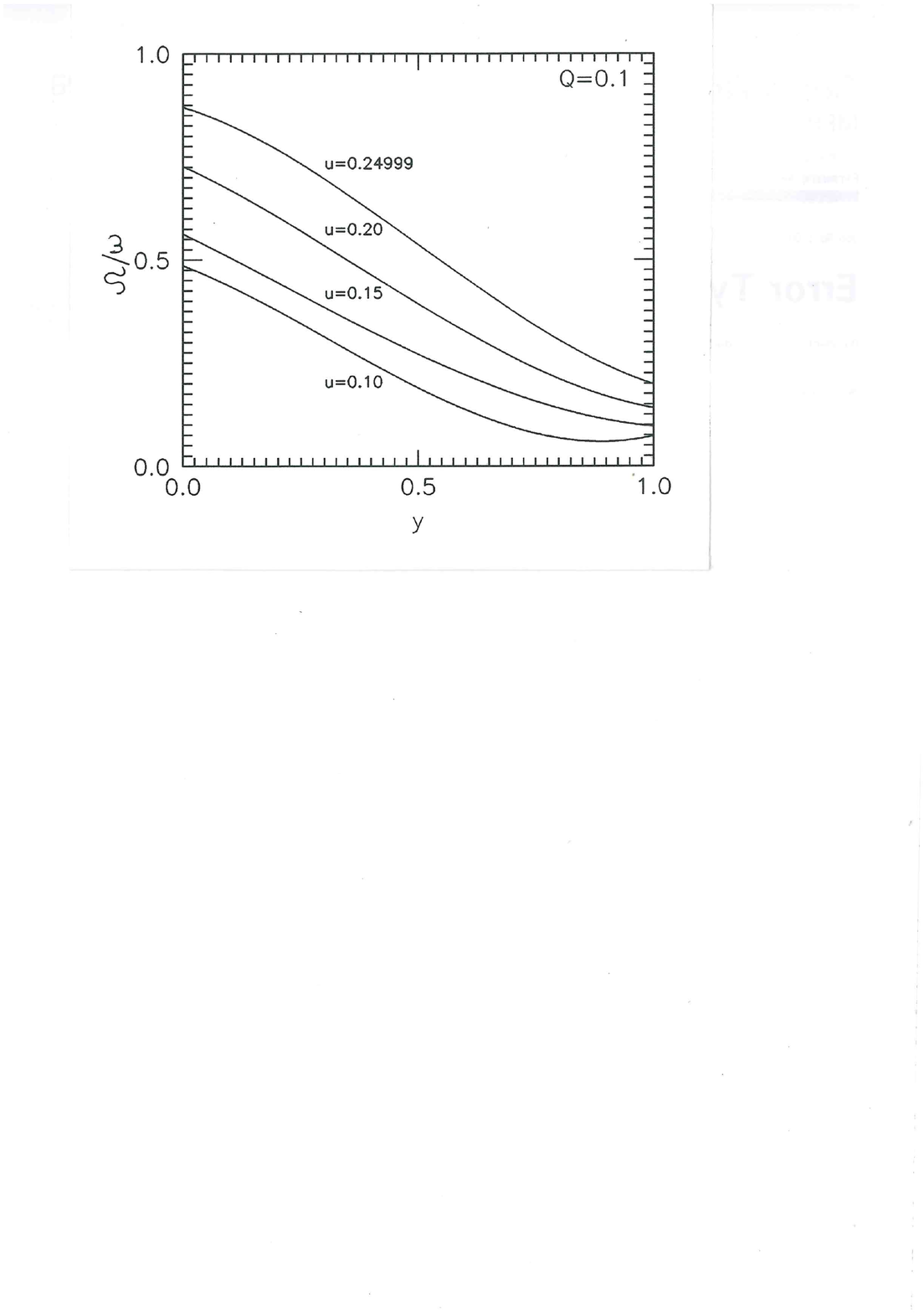}

\end{figure}







\begin{figure}
\caption{ The variation of relative drag ($\Omega/\omega$) vs. $y$, that is the change in drag from centre ($y=0$) to the surface ($y=1$) of the structures ($u =  0.10, 0.15, 0.20, 0.24999$) for the value of $Q = 0.3$.}
\includegraphics[angle=0,scale=1.0]{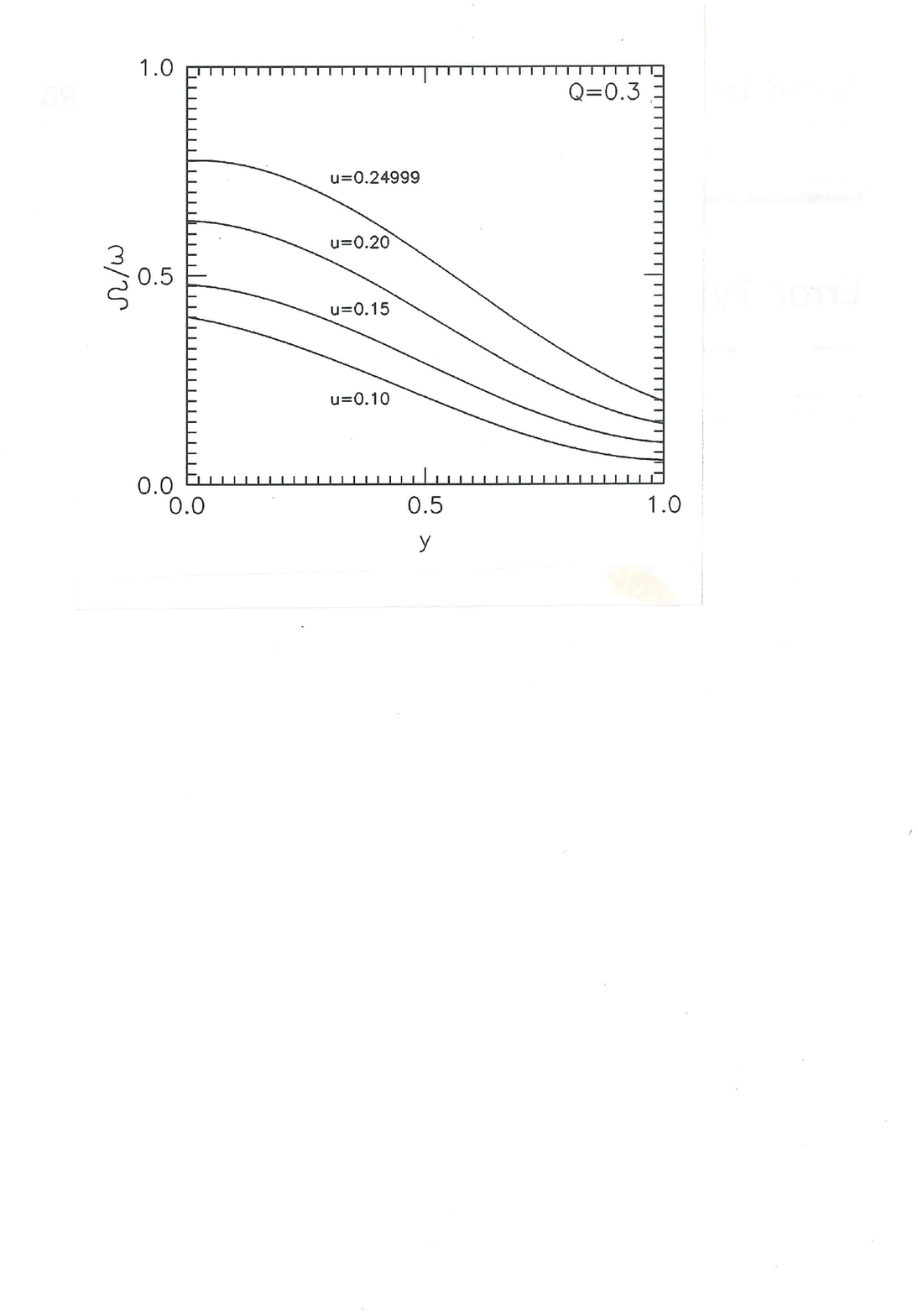}

\end{figure}

\begin{figure}
\caption{ The variation of relative drag ($\Omega/\omega$) vs. $y$, that is the change in drag from centre ($y=0$) to the surface ($y=1$) of the structures ($u =  0.10, 0.15, 0.20, 0.24999$) for the value of $Q = 0.6$.}
\includegraphics[angle=0,scale=1.0]{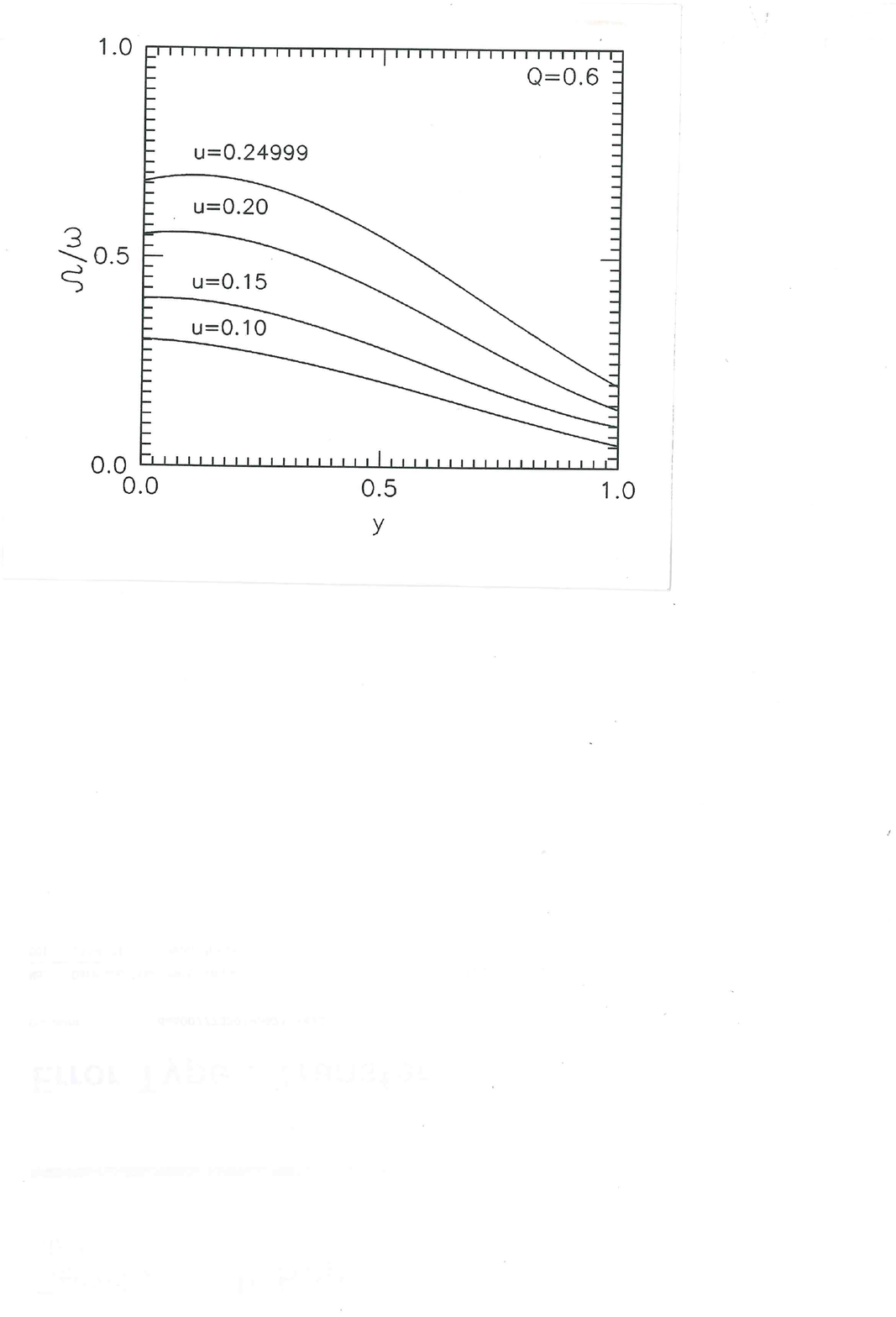}

\end{figure}
\begin{figure}
\caption{ The variation of relative drag ($\Omega/\omega$) vs. $y$, that is the change in drag from centre ($y=0$) to the surface ($y=1$) of the structures ($u =  0.10, 0.15, 0.20, 0.24999$) for the value of $Q = 0.9$.}
\includegraphics[angle=0,scale=1.0]{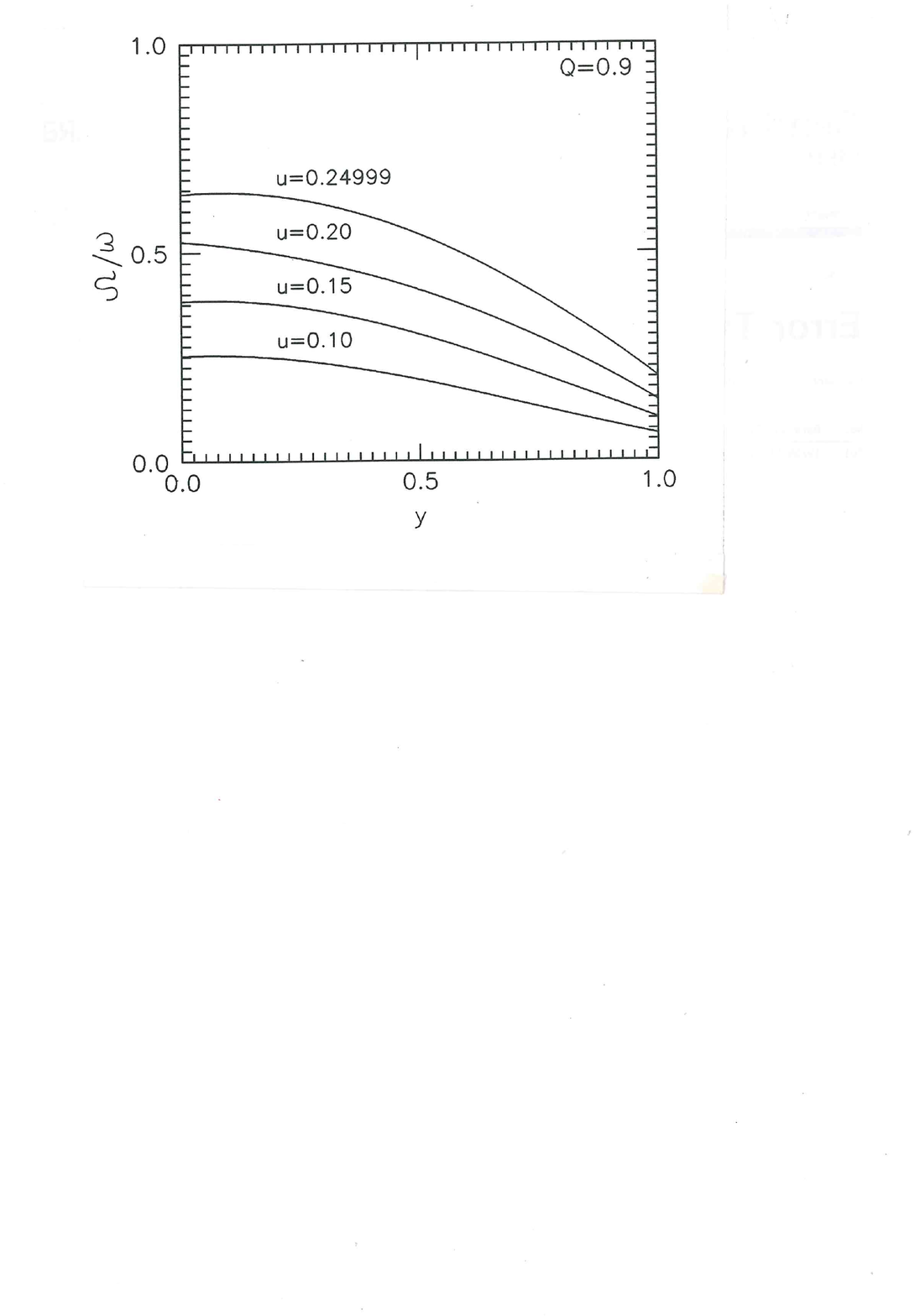}

\end{figure}

\begin{figure}

\caption{ The variation of relative drag ($\Omega/\omega$) vs. $y$, that is the change in drag from centre ($y=0$) to the surface ($y=1$) of the structures ($u =  0.10, 0.15, 0.20, 0.24999$) for the value of $Q = 1.1999$.}
\includegraphics[angle=0,scale=1.0]{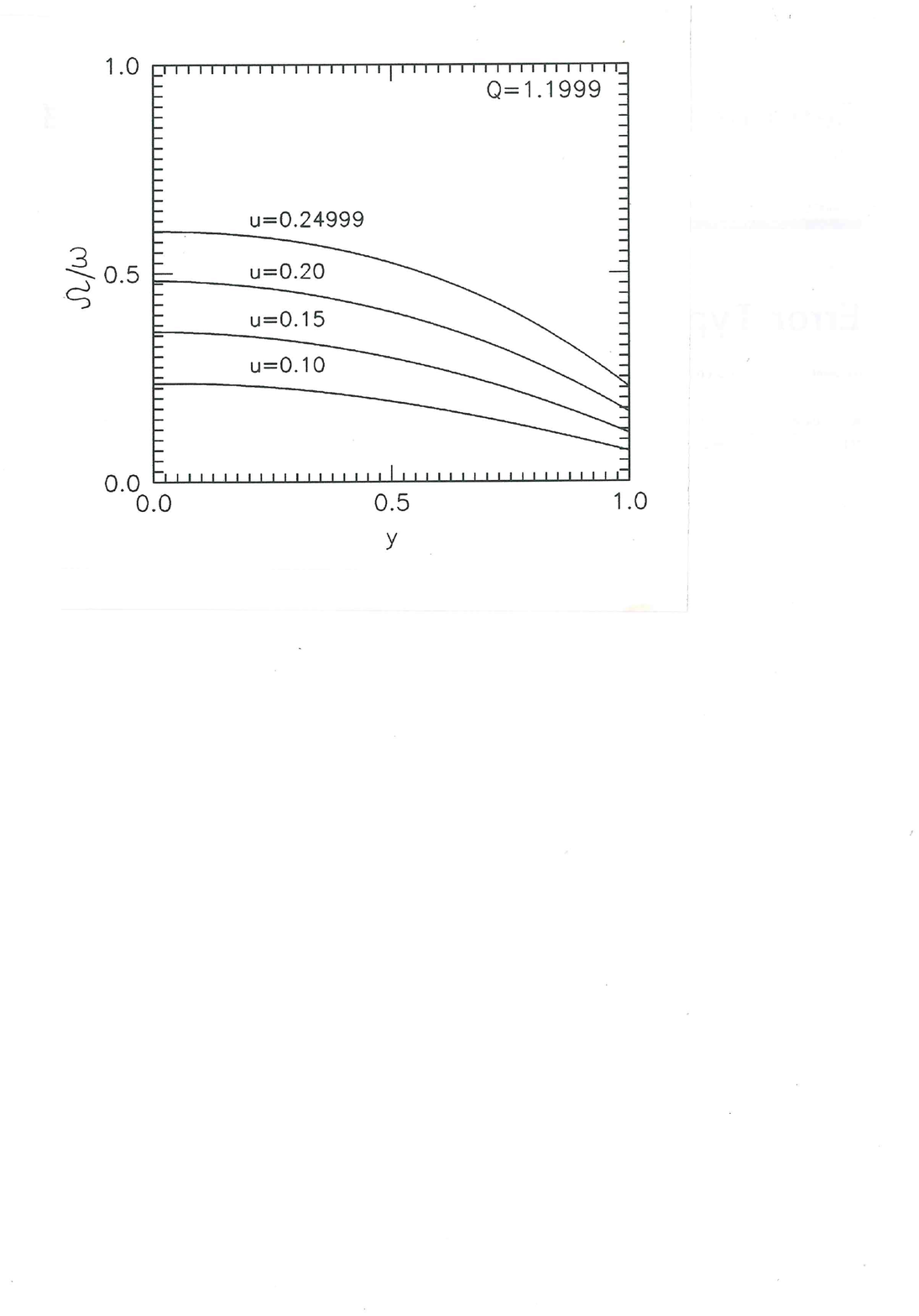}
\end{figure}

\clearpage

\begin{table}
\begin{center}
\caption{Moment of Inertia ($I_{\rm total}$), Mass ($M/M_\odot$), Size ($a$), Core size ($b$), Core-envelope boundary density $E_b ( 10^{14}\rm g{cm}^{-3})$, Central density $E_0 ( 10^{15}\rm g{cm}^{-3}$), Fractional moment of Inertia ($G_h = I_{\rm core}/I_{\rm total}$), and Surface red-shift ($z_a$) for the core-envelope models corresponding to an assigned value of surface density $E_a = 2\times 10^{14}\rm g{cm}^{-3}$ and for various values of  $u$ and $Q$ .\label{tbl-2}}
\begin{tabular}{crrrrrrrrrr}
\tableline\tableline
$I_{\rm total} \geq (\rm g{cm}^2)$ &$ u $& $Q$ & $(b/a)$ & $E_b$ & $ E_0$ & $a(\rm {km})$ & $b(\rm {km})$ & $M/M_\odot$ & $ (G_h)$ & $z_a$ \\
\tableline
\nodata &0.10250 &0.9 &0.866  & 2.7 & 1.6 &7.413 &6.420 &0.515  &0.650 &0.122\\
\nodata &0.10500 &0.6 &0.707 & 4.0 & 2.4 &7.503 &5.305  &0.534  &0.350 &0.125 \\
$1.8\times 10^{44}$&0.10875&0.3 &0.500 &8.0& 4.8 &7.636  &3.818 &0.563  &0.124 &0.130\\
\nodata &0.11125 &0.1 &0.289 &24.0&14.4&7.723 &2.232 &0.583  &0.024 &0.134\\
\tableline
\nodata &0.13375 &1.1999 &1.000  &2.0&1.2& 8.468 &8.468 &0.768  &1.000 &0.168\\
\nodata &0.13875 &0.9&0.866 &2.7&1.6& 8.625&7.469  &0.812  &0.650 &0.176 \\
$4\times 10^{44}$&0.14250 &0.6 &0.707 &4.0&2.4 &8.741 &6.180  &0.845 &0.352&0.183\\
\nodata &0.14625 &0.3 &0.500 &8.0&4.8 &8.855 &4.428 &0.878  &0.125 &0.189\\
\nodata &0.14875 &0.1 &0.289 &24.0&14.4 &8.930 &2.581 &0.901 &0.024 &0.193\\
\tableline
\nodata &0.23500 &1.1999 &1.000 &2.0&1.2 &11.225 &11.225 &1.790  &1.000 &0.374\\
\nodata &0.24250 &0.9&0.866& 2.7&1.6 &11.403&9.875  &1.876  &0.650 &0.393 \\
$2\times 10^{45}$&0.24625 &0.6 &0.707 &4.0&2.4  &11.490 &8.123 &1.920 &0.352&0.404\\
\nodata &0.25000 &0.3 &0.500 &8.0&4.8 &11.578 &5.789 &1.964  &0.125 &0.414\\
\tableline
\end{tabular}
\end{center}
\end{table}



\end{document}